\documentclass[a4paper,11pt]{article}

\usepackage{pos}
\usepackage{lipsum} 
\usepackage{caption}
\usepackage{subcaption}
\usepackage{lineno}

\title{Measuring the Astrophysical Galactic Plane Neutrino Flux and Searching for Galactic PeVatrons using the IceCube Multi-Flavor Astrophysical Neutrino Sample}

\ShortTitle{Multi-Flavor Neutrino Measurement of the Galactic Plane}

\author{The IceCube Collaboration \\{\normalsize \normalfont(a complete list of authors can be found at the end of the proceedings)}\\}

\emailAdd{thiesmeyer@icecube.wisc.edu}
\emailAdd{tyuan@icecube.wisc.edu}
\emailAdd{lseen@icecube.wisc.edu}
\emailAdd{lu.lu@icecube.wisc.edu}
\emailAdd{karle@icecube.wisc.edu}

\abstract{

Abstract: The IceCube Neutrino Observatory has provided new insights into the high-energy universe, in particular, unveiling neutrinos from the galactic plane. However, galactic neutrino sources are still unresolved. The recent detection of multi-PeV photons by LHAASO from the Cygnus region highlights its potential as a galactic neutrino source. Additionally, LHAASO, HAWC, and HESS have reported over forty galactic gamma-ray sources with energies above 100 TeV. Detecting neutrinos correlated with high-energy gamma-ray sources would provide compelling evidence of cosmic-ray acceleration in these galactic sources. In this work, we compile a 12.3-year, full-sky, all-flavor dataset, the IceCube Multi-Flavor Astrophysics Neutrino sample (ICEMAN). ICEMAN is the combination of three largely independent neutrino samples of different event morphologies and builds upon the previous work of the DNN-based cascade sample, Enhanced Starting Track Event Selection, and the Northern Track sample. Recent improvements in ice modeling and detector calibration are also incorporated into the cascade reconstruction. In addition to revisiting the galactic plane, we adopt two different analysis methods to search for galactic PeVatrons. First, we use a template-based approach to probe the Cygnus Cocoon region. Second, we use a point source hypothesis to find correlations between IceCube neutrinos and gamma-ray sources detected at energies greater than 100 TeV. 

\vspace{4mm}

{\bfseries Corresponding authors:}

Matthias Thiesmeyer$^{1*}$, 
Tianlu Yuan$^{1}$, 
Leo Seen$^{1}$, 
Lu Lu$^{1}$, 
Albrecht Karle$^{1}$\\

{$^{1}$ \itshape Dept. of Physics and Wisconsin IceCube Particle Astrophysics Center, University of Wisconsin—Madison}\\[4mm]
$^*$ Presenter
}

\ConferenceLogo{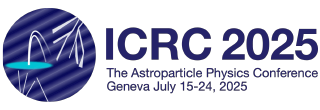}

\FullConference{39th International Cosmic Ray Conference (ICRC2025)\\
 15–24 July 2025\\
Geneva, Switzerland\\}

\begin{document}

\maketitle

\section{Introduction}\label{sec1}
The IceCube Neutrino Observatory is a cubic-kilometer neutrino detector located at the geographic south pole. In over ten years of operation time, IceCube has found evidence of multiple neutrino point sources, including the galaxy NGC 1068 \cite{NGC_1068_science} and the blazar TXS 0606+056 \cite{TXS_paper}. Both of these sources have been identified using \textit{track}-like signatures in the detector, coming from charged current muon neutrino interactions from the North.

Charged-current interactions of electron and tau neutrinos, as well as neutral-current interactions of all flavors can take the morphology of \textit{cascade}-like events in the detector. Reconstructed neutrinos from these interaction channels have been successfully used to observe the galactic plane as a neutrino source at 4.5$\sigma$ significance \cite{GP_science_paper}. Several other galactic plane analyses have been performed with different detection channels by IceCube \cite{ESETS_paper}\cite{NT_paper} and the ANTARES neutrino detector \cite{antares_gp}, yielding only mildly significant detections of the galactic plane.

In this analysis, we report a combined fit utilizing starting muon tracks, cascades, and through-going muon tracks from the north, which are combined into a single dataset---the IceCube Multi-Flavor Astrophysical Neutrino Sample (ICEMAN). We use ICEMAN to provide an updated measurement of the galactic plane as a neutrino source, with an expected median local significance of 5.5$\sigma$.

\section{Dataset}\label{sec2}
Three different event selections are combined to obtain an all-flavor, all-sky, combined dataset. The three different selections consist of cascades, starting tracks, and through-going tracks.
\begin{figure}[h]
     \centering
     \begin{subfigure}{0.47\textwidth}
         \centering
         \includegraphics[width=\linewidth]{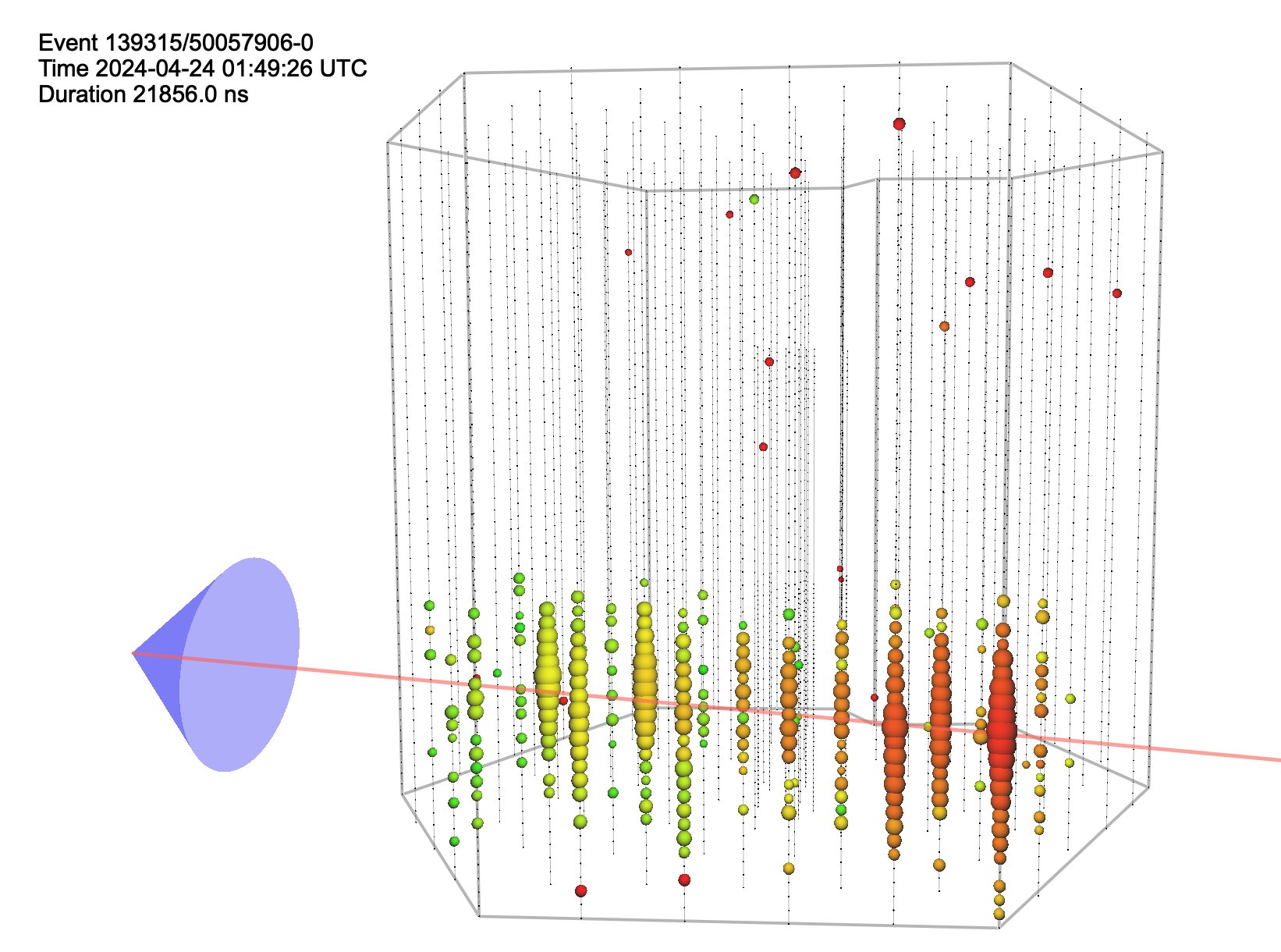}
         \caption{Example through-going track event}
         \label{fig:1a}
     \end{subfigure}
     \begin{subfigure}{0.47\textwidth}
         \centering
         \includegraphics[width=\linewidth]{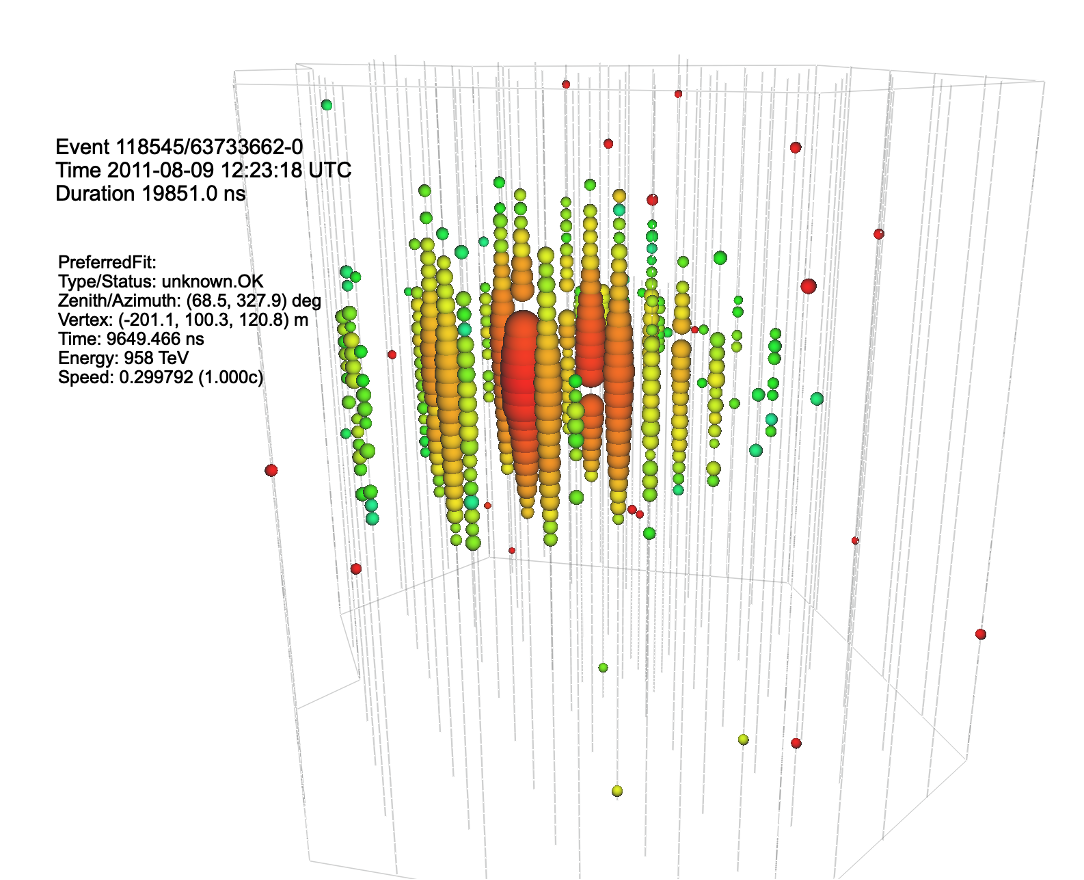}
         \caption{Example cascade event}
         \label{fig:1b}
     \end{subfigure}
     \caption{Two event views showing the two main topologies used in this analysis. A through-going track event (a) and a cascade event (b).}
     \label{fig:1}
\end{figure}
Cascade events are produced from neutral-current neutrino interactions of all flavors and charged-current interactions from electron and tau neutrinos. Due to their different morphology they can be more easily distinguished from track-like muons. This allows the rejection of the dominant background of atmospheric muons in the south. Therefore, this event selection does not rely on Earth absorption for background rejection and is applicable to the whole sky. Compared to track-like events, they have worse angular resolution but better energy resolution. At 10 TeV, cascades have a median angular resolution of approximately 10° compared to about 0.5° for through-going tracks. However, their median energy uncertainty at 10 TeV is 3\%, which is significantly better than the about 80\% median energy resolution of through-going tracks.
The cascade dataset used in this analysis uses a deep neural network (DNN) in the selection of cascade-like events, and is referred to as DNN cascades (DNNC). Various improvements \cite{IceMan_DNNC_proceeding} have been made to this dataset since the previous iteration \cite{GP_science_paper}. The new version of this dataset now includes 85,199 events detected between the 13th of May, 2011 to the 28th of November, 2023.

Through-going tracks from the north, Northern Tracks (NT), are mainly produced by charged-current interactions of muon neutrinos. Because the interaction vertex can be outside the detector, the effective area is very large. To filter out atmospheric muons via Earth absorption, only neutrino candidates with zenith angles above 85 degrees, i.e, a bit more than half the sky, are selected. This leads to over 99.8\% neutrino purity, i.e, a very low contamination of muons. The muon tracks induced from neutrino interactions have better angular resolution but a worse energy reconstruction compared to cascades. The dataset contains 13 years of IceCube data from the first of June, 2010 to the 28th of November, 2023. It has been used before in other IceCube point-source analyses \cite{13yNT_proceeding}.

The third subset consists of starting muon tracks. It is commonly referred to as the enhanced starting track event selection (ESTES). Like the through-going tracks from the north, these events are induced by charged current interactions of muon neutrinos. However, for starting tracks, the interaction vertex is inside the detector. This leads to a significant reduction of the atmospheric muon background, allowing for a full-sky application as well, at the cost of a smaller effective area. At 10 TeV, this data set has a median angular uncertainty of about one degree and a median energy uncertainty of about 30\%. The dataset contains data from the 13th of May 2011 to the 28th of November 2023. It contains a total of 11,755 events. A version of this dataset with 10 years of livetime has been used in previous IceCube publications \cite{ESETS_paper}.

\begin{figure}[h]
\centering
\includegraphics[width=0.6\linewidth]{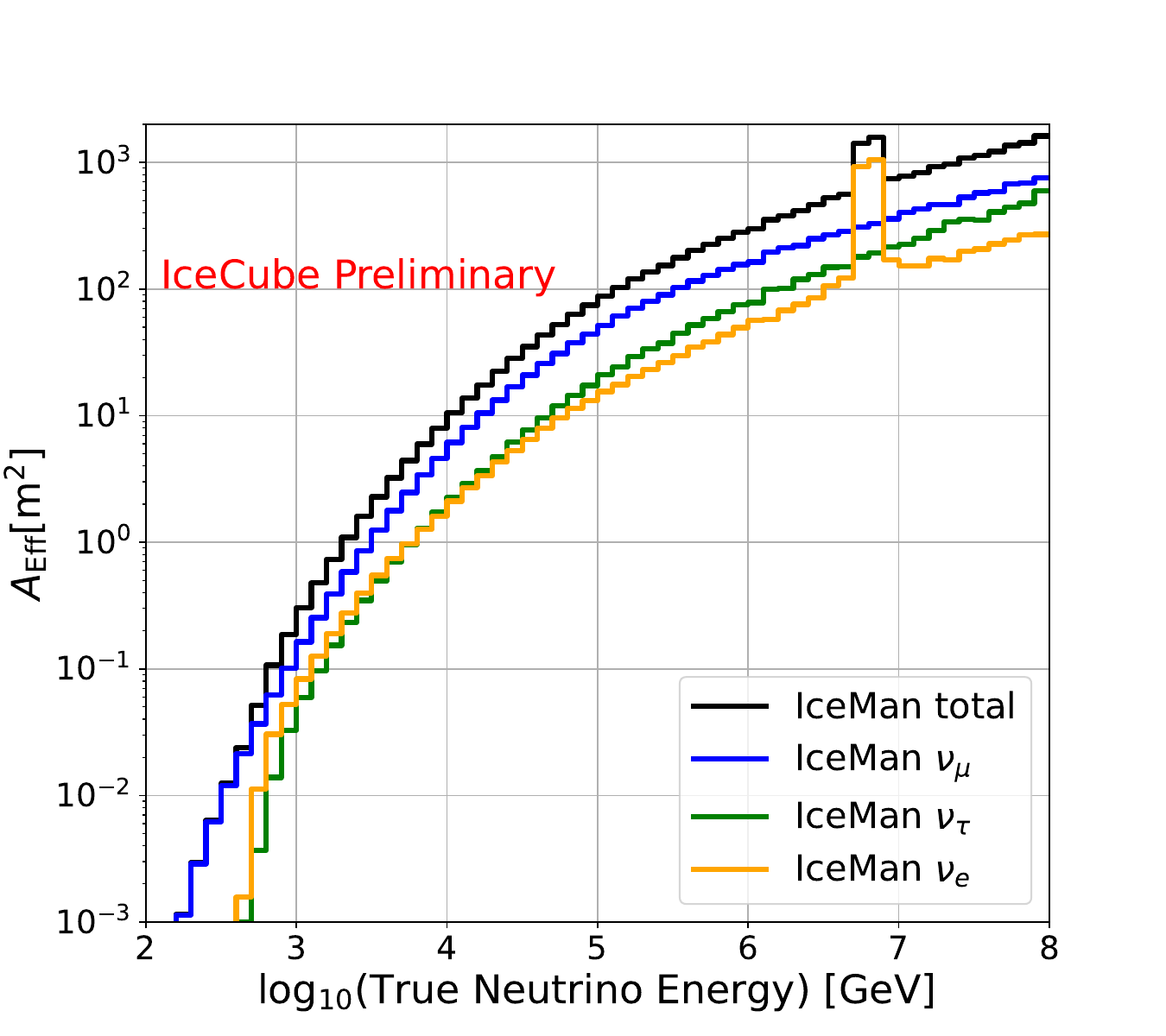}
\caption{Per-flavor all-sky effective area in the ICEMAN sample.}\label{fig:effective_area}
\end{figure}
\begin{figure}[h]
\centering
\includegraphics[width=0.6\linewidth]{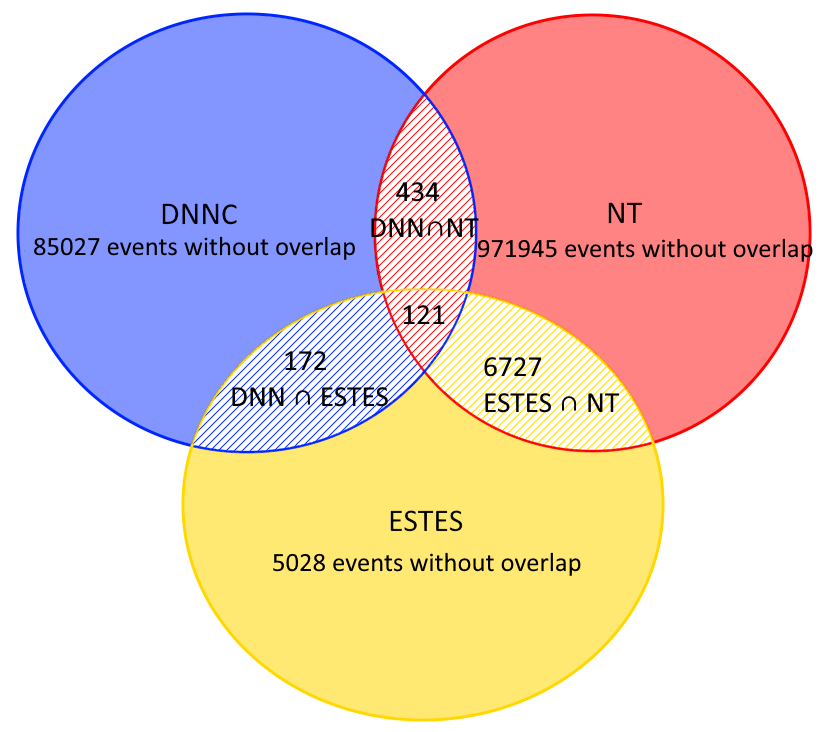}
\caption{Sketch showing which dataset overlapping events are kept in. The event numbers exclusive to the datasets and in all overlapping zones are shown. The colors of the shaded regions show which dataset the overlap will stay in.}\label{fig:overlaps}
\end{figure}

The per-flavor effective areas of the ICEMAN dataset are shown in Figure \ref{fig:effective_area}. In the combined sample, overlaps between the three different datasets are removed. The biggest set of overlaps is between ESTES and NT. These overlapping events are kept in the starting track set to maximize sensitivity. Figure \ref{fig:overlaps} shows the exact number of overlaps.
\section{Method}\label{sec3}
This is a template-based analysis where templates based on four different galactic emission models are being tested. The Fermi $\pi^0$ template \cite{Fermi_paper}, the KRA$_{\gamma}^5$ and KRA$_{\gamma}^{50}$ templates \cite{Kra_paper} and the CRINGE template \cite{CRINGE_paper} with an unresolved sources component \cite{unresolved_model}. The results of this analysis will include the measurement of the model normalization for each of the four model hypotheses, the rejection of the null hypothesis under each model assumption, and the rejection of the global null hypothesis by the most significant model hypothesis after trial correction. We use an unbinned maximum likelihood method that utilizes the reconstructed direction, energy, and angular uncertainty of each neutrino candidate in the ICEMAN dataset.

The signal considered is the excess of neutrinos along the galactic plane over the only declination-dependent background fluxes of atmospheric muons, atmospheric neutrinos, and diffuse astrophysical neutrinos.

\begin{figure}[h]
\centering
\includegraphics[width=0.9\linewidth]{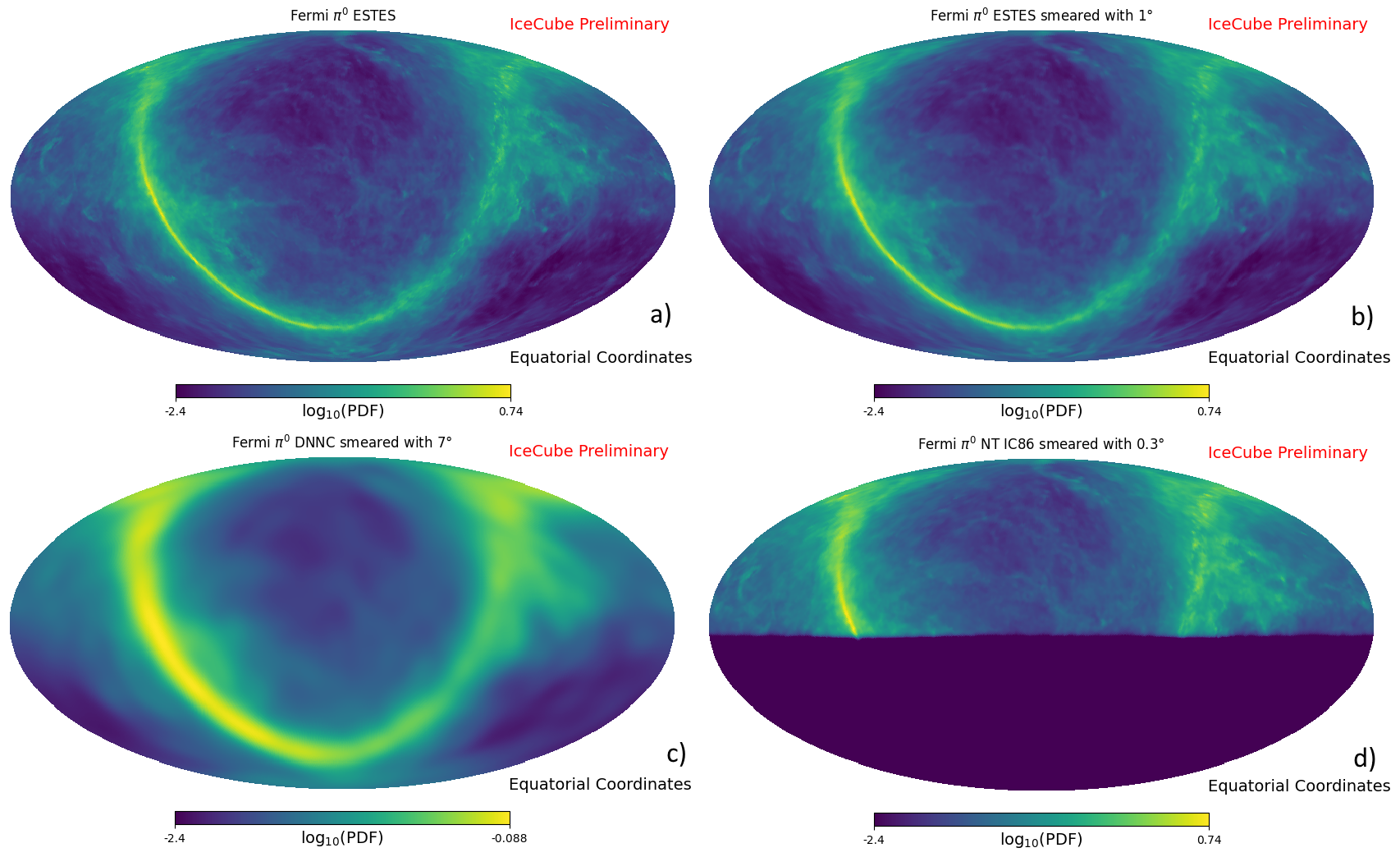}
\caption{Spatial signal PDF for the Fermi $\pi^0$ template. Shown in a) acceptance weighted with ESTES without smearing. In b) acceptance weighted with ESTES and smeared with a Gaussian kernel of 0.3 degrees width. In c) the template is acceptance weighted with DNN Cascades and smeared with a Gaussian kernel of 7 degrees width. In d) the template is acceptance weighted with the Northern Tracks and smeared with a Gaussian of 0.3 degree width. These smearing widths are chosen based on the median angular uncertainty of the signal events in the respective datasets.}\label{fig:pdf_pi0_estes}
\end{figure}

To parameterize this excess prediction based on each model, templates of their respective prediction are folded with the effective area of each dataset. After normalizing, this creates a spatial probability density function (PDF)  for each dataset and template combination. To account for the respective angular uncertainty of each event, each spatial PDF is smeared with a set of 2D Gaussian kernels corresponding to different angular uncertainties. They are then evaluated based on the uncertainty of the respective event. Figure \ref{fig:pdf_pi0_estes} shows an example of the unsmeared PDF as well as an example of the smeared PDF for each sub-dataset.

\begin{figure}[h]
\centering
\includegraphics[width=0.5\linewidth]{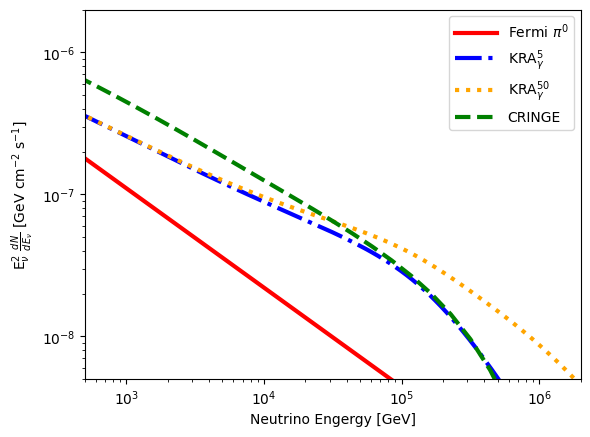}
\caption{Nominal per-flavor flux prediction for all four templates used in this analysis.}\label{fig:nominal_flux}
\end{figure}

To build the energy PDF, the MC is weighted with a single power law with a spectral index of -2.7 is used for the Fermi $\pi^0$ template. For the other three templates, the sky-averaged spectrum predicted by each template is used. The weighted MC is then binned in reconstructed energy. This way an energy PDF in reconstructed energy based on the predicted energy spectra is obtained. All nominal all-sky fluxes are shown in Figure \ref{fig:nominal_flux}.

To parameterize the declination-dependent backgrounds, a PDF of the density of neutrinos per declination is used. Since this background PDF is derived from data that contains partial contamination from the signal, a signal subtraction method is applied. The observed PDF from data $\bar{D}$, is characterized as the sum of the true isotropic background $B$, and the average density of the signal per declination band $\bar{S}(\delta_i)$:
\begin{equation}\label{eq:sigsub}
    \bar{D}(\delta_i, E_i) = \frac{n_s}{N} \bar{S}(\delta_i, E_i) + (1-\frac{n_s}{N}) B(\delta_i, E_i).
\end{equation}
Using this, the likelihood of observing $n_s$ neutrinos is defined as shown in equation \ref{eq:llh}:
\begin{equation}\label{eq:llh}
\mathcal{L}(n_s) = \prod_{i=1}^N 
\frac{n_s}{N}
S(\delta_i, \alpha_i, E_i, \sigma_i) + 
\bar{D}_i(\sin(\delta_i), E_i) - 
\frac{n_s}{N} \bar{S}_i (\sin(\delta_i), E_i).
\end{equation}
Then, the test statistic (TS) is defined as the likelihood ratio of fitting $n_s$  neutrinos over zero neutrinos,
\begin{equation}\label{eq:TS}
\mathrm{TS} = 2 \ln \left[ \frac{\mathcal{L}(\hat{n}_s)}{\mathcal{L}(n_s = 0)} \right].
\end{equation}

Using this method, we can derive the sensitivity and discovery potential of this analysis using pseudo-experiments.
\begin{table}[h]
    \centering
    \begin{tabular}{l|c|c|c|c}
        \hline
        Quantity & Fermi $\pi^0$ [10$^{-12}$ TeV cm$^{-2}$ s$^{-1}$]& KRA$_{\gamma}^{5}$& KRA$_{\gamma}^{50}$ & CRINGE\\
        \hline
        Sensitivity &  4.68 & 0.13 & 0.10 & 0.13\\
        \hline
        Discovery Potential & 18.7 & 0.49 & 0.40 & 0.53\\
        \hline
    \end{tabular}
    \caption{Sensitivity and discovery potentials for different models. For Fermi $\pi^0$, the per-flavor flux at 100 TeV is reported. For the other three models, sensitivity and discovery potential are reported in units of model flux.}
    \label{tab:sensitivity_discovery_potential}
\end{table}
Here, a number of Monte-Carlo events corresponding to an injected flux are sampled from the spatial and energy distribution predicted by a chosen model. Data, with the right ascension coordinate randomized, is added to these signal events to account for the isotropic backgrounds. Table \ref{tab:sensitivity_discovery_potential} shows the necessary flux for 50$\%$ of pseudo experiments to exceed the TS corresponding to a 5$\sigma$ discovery for the four different templates. Since the flux measured in \cite{GP_science_paper} is above the flux necessary to reach the 5$\sigma$ threshold for the Fermi $\pi^0$ model, a 5$\sigma$ pre-trial significance is a likely outcome of this analysis. In a set of pseudo-experiments, injecting the best fit flux of the previous measurement, we obtain a median local significance of 5.5$\sigma$ for the Fermi $\pi^0$ model, as shown in Table \ref{tab:expected_significance}.
\begin{table}[h]
    \centering
    \begin{tabular}{|c|c|c|c|c|}
        \hline
        Template & Fermi $\pi^0$ & KRA$_{\gamma}^{5}$& KRA$_{\gamma}^{50}$ & CRINGE \\ \hline
        Median Significance (injecting Fermi $\pi^0$) & 5.5$\sigma$ & 4.71$\sigma$ & 4.46$\sigma$ & 5.28$\sigma$ \\ \hline
    \end{tabular}
    \caption{Expected local significance from injecting the best fit model and flux of the previous analysis. }
    \label{tab:expected_significance}
\end{table}
\section{Conclusion}\label{sec5}
In this work, three different sub-datasets in IceCube have been combined into a single multi-flavor all-sky dataset. We have detailed the improvements and dataset combination methods. Furthermore, we have shown the measurement technique with a combined likelihood in space and energy. Finally, we have calculated the sensitivities and discovery potentials under all four different model hypotheses. When compared to the values from the previous analysis \cite{GP_science_paper} the sensitivities have improved by over 20\%. These improvements highlight the capabilities of the combined ICEMAN dataset to make an improved measurement of the galactic plane as an extended neutrino source. Future IceCube analyses will utilize the ICEMAN dataset to measure the neutrino flux from the Cygnus region as an extended source and search for galactic neutrino sources based on a multi-messenger informed catalog \cite{IceMan_DNNC_proceeding}.

\bibliographystyle{ICRC}
\bibliography{references}

\clearpage

\section*{Full Author List: IceCube Collaboration}

\scriptsize
\noindent
R. Abbasi$^{16}$,
M. Ackermann$^{63}$,
J. Adams$^{17}$,
S. K. Agarwalla$^{39,\: {\rm a}}$,
J. A. Aguilar$^{10}$,
M. Ahlers$^{21}$,
J.M. Alameddine$^{22}$,
S. Ali$^{35}$,
N. M. Amin$^{43}$,
K. Andeen$^{41}$
% ...
\\
\\
% ...
$^{10}$ Universit{\'e} Libre de Bruxelles, Science Faculty CP230, B-1050 Brussels, Belgium \\
% ...
$^{16}$ Department of Physics, Loyola University Chicago, Chicago, IL 60660, USA \\
$^{17}$ Dept. of Physics and Astronomy, University of Canterbury, Private Bag 4800, Christchurch, New Zealand \\
% ...
$^{21}$ Niels Bohr Institute, University of Copenhagen, DK-2100 Copenhagen, Denmark \\
$^{22}$ Dept. of Physics, TU Dortmund University, D-44221 Dortmund, Germany \\
% ...
$^{35}$ Dept. of Physics and Astronomy, University of Kansas, Lawrence, KS 66045, USA \\
% ...
$^{39}$ Dept. of Physics and Wisconsin IceCube Particle Astrophysics Center, University of Wisconsin{\textemdash}Madison, Madison, WI 53706, USA \\
% ...
$^{41}$ Department of Physics, Marquette University, Milwaukee, WI 53201, USA \\
% ...
$^{43}$ Bartol Research Institute and Dept. of Physics and Astronomy, University of Delaware, Newark, DE 19716, USA \\
% ...
$^{63}$ Deutsches Elektronen-Synchrotron DESY, Platanenallee 6, D-15738 Zeuthen, Germany \\
$^{\rm a}$ also at Institute of Physics, Sachivalaya Marg, Sainik School Post, Bhubaneswar 751005, India \\
% ...

\subsection*{Acknowledgments}

\noindent
The authors gratefully acknowledge the support from the following agencies and institutions:
USA {\textendash} U.S. National Science Foundation-Office of Polar Programs,
U.S. National Science Foundation-Physics Division,
U.S. National Science Foundation-EPSCoR,
U.S. National Science Foundation-Office of Advanced Cyberinfrastructure,
Wisconsin Alumni Research Foundation,
Center for High Throughput Computing (CHTC) at the University of Wisconsin{\textendash}Madison,
Open Science Grid (OSG),
Partnership to Advance Throughput Computing (PATh),
Advanced Cyberinfrastructure Coordination Ecosystem: Services {\&} Support (ACCESS),
Frontera and Ranch computing project at the Texas Advanced Computing Center,
U.S. Department of Energy-National Energy Research Scientific Computing Center,
Particle astrophysics research computing center at the University of Maryland,
Institute for Cyber-Enabled Research at Michigan State University,
Astroparticle physics computational facility at Marquette University,
NVIDIA Corporation,
and Google Cloud Platform;
Belgium {\textendash} Funds for Scientific Research (FRS-FNRS and FWO),
FWO Odysseus and Big Science programmes,
and Belgian Federal Science Policy Office (Belspo);
Germany {\textendash} Bundesministerium f{\"u}r Forschung, Technologie und Raumfahrt (BMFTR),
Deutsche Forschungsgemeinschaft (DFG),
Helmholtz Alliance for Astroparticle Physics (HAP),
Initiative and Networking Fund of the Helmholtz Association,
Deutsches Elektronen Synchrotron (DESY),
and High Performance Computing cluster of the RWTH Aachen;
Sweden {\textendash} Swedish Research Council,
Swedish Polar Research Secretariat,
Swedish National Infrastructure for Computing (SNIC),
and Knut and Alice Wallenberg Foundation;
European Union {\textendash} EGI Advanced Computing for research;
Australia {\textendash} Australian Research Council;
Canada {\textendash} Natural Sciences and Engineering Research Council of Canada,
Calcul Qu{\'e}bec, Compute Ontario, Canada Foundation for Innovation, WestGrid, and Digital Research Alliance of Canada;
Denmark {\textendash} Villum Fonden, Carlsberg Foundation, and European Commission;
New Zealand {\textendash} Marsden Fund;
Japan {\textendash} Japan Society for Promotion of Science (JSPS)
and Institute for Global Prominent Research (IGPR) of Chiba University;
Korea {\textendash} National Research Foundation of Korea (NRF);
Switzerland {\textendash} Swiss National Science Foundation (SNSF).

\end{document}